# AlGaN/AlN Stranski-Krastanov quantum dots for highly efficient electron beam pumped emitters: The role of miniaturization and composition to attain far UV-C emission

Jesus Cañas[*,1], Anjali Harikumar[1], Stephen T. Purcell[3], Nevine Rochat[2], Adeline Grenier[2], Audrey Jannaud[2], Edith Bellet-Amalric[1], Fabrice Donatini[4], and Eva Monroy[1]

[1] Univ. Grenoble-Alpes, CEA, Grenoble INP, IRIG, PHELIQS, 38000 Grenoble, France
[2] Univ. Grenoble Alpes, CEA, LETI, 38000 Grenoble, France
[3] Institut Lumière Matière, CNRS, University of Lyon, Univ. Claude Bernard Lyon 1, 69622 Villeurbanne, France
[4] Univ. Grenoble Alpes, CNRS, Grenoble INP, Institut Neel, 38000 Grenoble, France

* Corresponding author: jesus.canasfernandez@cea.fr

OrcID:
Jesus Cañas: 0000-0003-0202-6987
Anjali Harikumar: 0000-0002-0354-2142
Stephen T. Purcell: 0009-0000-3656-4703
Névine Rochat: 0000-0003-3574-4424
Edith Bellet-Amarlic: 0000-0003-2977-1725
Fabrice Donatini: 0000-0001-6556-1683
Eva Monroy: 0000-0001-5481-3267

**ABSTRACT:**

Conventional ultraviolet (UV) lamps for disinfection emit radiation in the 255-270 nm range, which poses a high risk of causing cancer and cataracts. To address these concerns, solid-state far UV-C sources emitting below 240 nm are gaining attention as a safe and sustainable disinfection solution for occupied spaces. Here, we delve into the extension of the $Al_xGa_{1-x}N$/AlN quantum dot (QD) technology towards the far UV-C range, which presents various challenges associated with the reduction of the lattice mismatch and band offset when Al is incorporated in the QDs. We explore the structural and optical impact of increasing the Al content through the increase of the Al flux and eventual correction of the Ga flux to maintain a constant metal/N ratio. We also examine the impact of extreme miniaturization of the QDs, achieved through a reduction of their growth time, on the spectral behavior and internal quantum efficiency ($IQE$). The high Al content results in QDs with a reduced aspect ratio (height/diameter) and thicker wetting layer when compared to the GaN/AlN system. Self-assembled QDs grown with a metal/N ratio ranging from 0.5 to 0.8 show an $IQE$ around 50%, independent of the Al content (up to 65%) or emission wavelength (300-230 nm). However, samples emitting at wavelengths below 270 nm exhibit a bimodal luminescence associated with inhomogeneous in-plane emission attributed to fluctuations of the QD shape associated with extended defects. Reducing the QD size exacerbates the bimodality without reducing the emission wavelength. The power efficiencies under electron beam pumping range from 0.4% to 1%, with clear potential for improvement through surface treatments that enhance light extraction efficiency.

# 1. INTRODUCTION

The fabrication of solid-state far UV-C sources emitting below 240 nm has motivated an important research effort in the past few years.[1] This research has been driven firstly by the demonstration that this radiation also possesses the strong germicidal properties of the traditional 256-270 nm range. Secondly, whereas traditional disinfection lamps are strongly carcinogenic and cataractogenic, far UV-C radiation does not penetrate human skin or eyes, making it a safer option for use in occupied spaces.[1–3]

The development of highly efficient light emitting diodes (LEDs), with blue LEDs now reaching close to 80% power efficiency,[4] has enabled significant reductions in energy consumption and facilitated the transition towards more sustainable technologies. However, in the UV range, the efficiency of LEDs declines drastically for emission wavelengths below 350 nm.[5] This is a consequence of the transition from the well-established InGaN semiconductor system to the promising but still immature AlGaN system. Some of the main technological difficulties involving AlGaN LEDs relate to their wide band gap, which causes inefficient p-type doping and resistive ohmic contacts. Despite these challenges, the development of AlGaN LEDs has been a focal point due to the high demand for UV emitters for disinfection,[5–7] and the need to replace highly toxic mercury lamps.

An alternative technology to AlGaN LEDs is the use of electron beam pumped lamps,[8] which consist of a UV-emitting material (anode) mounted in a vacuum tube. In this tube, a cold cathode pumps high-energy electrons into the anode, generating carriers through impact ionization. Unlike traditional AlGaN LEDs, this approach has the benefit of not needing p-type doping or contacts, as carrier generation is achieved through electron pumping in a large active volume (active region depth > 100 nm). Electron beam pumped lamps have the potential to overcome some of the technological barriers associated with AlGaN LEDs, and show promise for UV lighting applications, particularly in the UV-C below 240 nm, where demonstrated LED efficiencies are under 1%.

Electron-pumped devices based on AlGaN quantum wells (QWs) were first reported in 2010.[9] Since then, there have been remarkable advances in miniaturization,[10] efficiency improvements,[11] and output power,[12] with devices now capable of reaching watt-level outputs. The use of AlGaN quantum dots (QDs) represents a promising avenue for further improvement in device performance. The three-dimensional (3D) carrier confinement in QDs prevents the carrier diffusion to non-radiative centers and leads to an enhancement of the radiative efficiency at high temperature.[13] Plasma-assisted molecular beam epitaxy (PAMBE) has proven to be a highly effective technique for growing high-quality AlGaN QD superlattices. The Stranski–Krastanov growth mode that occurs in nitrogen-rich conditions enables the formation of high-density layers of small QDs,[13,14] fulfilling the requirements for UV emission. Previous reports showed that Stransky-Krastranov GaN/AlN QDs grown by PAMBE present photoluminescence decay times that do not depend on temperature in the range between 5 K and 300 K,[13] which confirms the suppression of nonradiative processes. UV-emitting $Al_xGa_{1-x}N$/AlN QDs with x ≤ 0.1 can attain an internal quantum efficiency (*IQE*) in the range of 10% to 40%,[15] higher than those generally obtained for $Al_xGa_{1-x}N$/$Al_yGa_{1-y}N$ QDs.[16] Further studies demonstrated that the *IQE* could be higher than 50% in the 276–296 nm spectral range in $Al_xGa_{1-x}N$/AlN QDs generated from the deposition of 4 to 5 monolayers (ML) of $Al_xGa_{1-x}N$ (always with x ≤ 0.1).[17] More recent studies showed that even smaller (2.5-3 ML) $Al_xGa_{1-x}N$/AlN QDs (x ≤ 0.1) could attain *IQE*s close to 50% if the metal/N ratio during the growth remained around 0.4, dropping for smaller metal/N values.[18] Furthermore, their efficiencies remain stable as a function of the optical pumping power up to 200 kW/cm²,[18] proving that such QDs can be used for application in devices with various pumping requirements. Overall, the incorporation of AlGaN QD layers in UV has garnered considerable attention, showcasing promising outcomes for the integration of QDs into devices that emitting in the first disinfection band, around 270 nm.[19–22] However, extending this technology to the far UV-C range presents additional challenges due to the higher Al contents needed in the QDs such as (i) the impact of the reduced lattice mismatch on the QD morphology, since the QD formation is not driven by the accumulated elastic energy in the layer but by the low surface energy with respect to the basal (0001) plane,[23] (ii) the effect of the reduced band offset on the carrier confinement, (iii) the enhanced density of point defects in the Al containing alloy (vacancies and pollutants), and (iv) the structural defects generated by the necessity to grow at lower temperatures than ideal for AlN due to the significant difference in binding energy between Al-N and Ga-N.

In this work, we explore the extension of the $Al_xGa_{1-x}N$/AlN QD technology towards the far UV-C range. We examine the structural and optical impact of increasing the aluminum content in the QDs through the incorporation of an additional Al flux and eventual correction of the Ga flux to maintain a constant metal/N ratio. Furthermore, we explore the effect of extreme miniaturization of the QDs on the spectral behavior and internal quantum efficiency.

## 2. EXPERIMENTAL SECTION

The samples under study consist of 100 periods of $Al_xGa_{x-1}N$/AlN self-assembled QDs deposited on commercial 1-μm-thick (0001)-oriented AlN-on-sapphire templates by PAMBE. The substrates were degreased and mounted with indium on a molybdenum sample holder. During the growth, the substrate temperature was fixed at 720°C, and the active nitrogen flux was adjusted to provide a growth rate of 0.6 monolayers per second (ML/s) under metal-rich conditions. For the generation of $Al_xGa_{x-1}N$ QDs, the metal/N flux ratio was kept below the stoichiometric value. This resulted in the Stranski-Krastanov growth mode due to the lattice mismatch and the energetically favorable {10-13} facets over the (0001) plane under N-rich conditions.[23] The growth process was monitored by reflection high-energy electron diffraction (RHEED). The $Al_xGa_{x-1}N$ deposition time was set between 12 s and 15 s, during which roughening was observed in the RHEED pattern. The AlN overgrowth to form the barriers was performed under Al-rich conditions (Al/N ≈ 1.1), to favor the planarization of the surface. After growing 4 nm of AlN, the Al cell was shuttered and the excess of metal was consumed with nitrogen. A schematic description of the complete structures is presented in figure 1a.

The structural properties of the samples were studied by scanning transmission electron microscopy (STEM) in bright field (BF) and high-angle annular dark field (HAADF). The observations were performed on an aberration-corrected FEI Titan 80-300 microscope operated at 200 kV. The periodicity and structural quality were analyzed by x-ray diffraction (XRD) in a Rigaku SmartLab diffractometer using a 2 bounce Ge(220) monochromator and a long plate collimator of 0.228° for the secondary optics.

Cathodoluminescence was performed in a FEI Inspect F50 field-emission scanning electron microscope (SEM) equipped with a liquid-helium-cooled Gatan stage. The experiments conducted to record the emission spectra and extract the internal quantum efficiency were performed using 50×50 μm$^2$ scans with an acceleration voltage of 5 kV, injecting currents around 100 pA. The CL signal was collected through a parabolic mirror and analyzed with a 550 mm focal length iHR550 spectrometer equipped with 600 grooves/mm diffraction grating. The spectra were recorded with a thermoelectric cooled silicon Andor Technology Newton DU940 BU2 CCD camera.

Additional CL experiments to extract λ-filtered maps were carried out using an Attolight CL microscope. The acceleration voltage was 10 kV and the beam current was ≈ 5 nA. The luminescence was collected through an integrated microscope objective (numerical aperture =

0.7). By scanning the sample, the optical spectra of each pixel are recorded on a CCD camera through a dispersive spectrometer with 400 mm focal length (grating: 150 grooves/mm blazed at 500nm).

Finally, the angular distribution of the CL to estimate the power efficiency was performed in a CL setup where the electron beam excitation was provided by a Kimball Physics EGPS-3212 electron gun operated in direct current mode, under normal incidence. The beam spot diameter was 1-1.5 mm. The gun was operated with an acceleration voltage of 5-6 kV. The CL emission was analyzed using a calibrated GaP UV photodetector attached to a rotator with a circular motion around the sample. It was verified that the luminescence instensity increased linearly with the injected current in the range of 1 nA to 500 µA.

The electronic structure of the QDs was modelled using the Nextnano3 8-band k·p Schrödinger-Poisson equation solver[24] with the GaN and AlN parameters described by Kandaswamy, et al.[25]. For $Al_xGa_{1-x}N$ alloys, all the bowing parameters were set to zero. For the calculation of the band diagram, the spontaneous and piezoelectric polarization and the band gap deformation potentials were taken into account.

## 3. RESULTS AND DISCUSSION

The growth of $Al_xGa_{x-1}N$ QD superlattices presents multiple degrees of freedom concerning the parameters used during the epitaxy. The choice of these parameters (e.g. III/V flux ratio, Al/Ga ratio or deposition time) is of paramount importance as it can have dramatic effects on the emission wavelength and efficiency. Aiming for the shortest wavelength (far UV-C) with the maximum efficiency requires thorough optimization to obtain small enough nanostructures with high Al content, while keeping the advantages of 3D carrier confinement.

In order to understand the influence of the growth parameters on the emission properties in the UV range, three series of AlGaN/AlN QD samples with different conditions for the formation of the QDs were investigated: a first series where the Ga flux was kept constant and the Al flux increased during the AlGaN deposition, hence varying the III/V flux ratio, a second series where the Al/Ga flux ratio was modified while keeping a constant metal/N flux ratio, and a third series where the QD deposition time was reduced.

**Series 1: Constant Ga flux**

A first set of samples (S1-S5) is dedicated to the investigation of the incorporation of an aluminum flux ($\Phi_{Al}$, varying from 0 to 0.2 ML/s) while keeping constant the flux of Ga ($\Phi_{Ga}$ = 0.3 ML/s) and active nitrogen ($\Phi_N$ = 0.6 ML/s), to increase the emission energy of the QDs. The addition of $\Phi_{Al}$ implies an increase of the metal/N flux ratio, $(\Phi_{Al} + \Phi_{Ga})/\Phi_N$, during the QD deposition. In ref. [18], we demonstrated that it is possible to grow $Al_xGa_{1-x}N$ QDs (x = 0, 0.1) with high $IQE$ (around 50%) when keeping the metal/N ratio below 0.75. Under these conditions, the growth kinetics are dominated by Ga, which is highly mobile on the growing surface at the QD growth temperature. In contrast, the mobility of Al is negligible at such growth temperature, which can have an impact on the geometry and optical performance of QDs with high Al content.

The growth parameters, nominal AlGaN composition, and structural and optical characterization of the samples in series 1 are summarized in table 1. In the reference sample, S1, the deposition of pure GaN during 15 s (4.5 ML) leads to the spontaneous formation of GaN QDs on AlN. A θ−2θ XRD scan around the (0002) reflection of AlN, showing a superlattice period of 5.5 nm, is presented as Supporting Information. For this sample, the CL emission is located at 314 nm at 5 K and at 318 nm at room temperature. The comparison between the integrated CL intensity at room temperature and at 5 K allows an estimation of the $IQE$,[18] which yields a value of 54% (average of measurements in six different areas of the sample), comparable with previous results.[18] The emission inhomogeneity of the sample surface is assessed using the standard deviation of the $IQE$, represented as $\Delta(IQE)$, on the studied areas of the sample. The sample S1 presents a very homogeneous emission with $\frac{\Delta IQE}{IQE} < 10\%$, i.e. $IQE \approx (54\pm5)\%$.

Upon increasing $\Phi_{Al}$ to 0.2 ML/s, the nominal Al concentration in the QDs rises to 40% for sample S5, and the metal/N ratio evolves from 0.50 to 0.83, i.e. the nitrogen excess is reduced as the series progresses. As the deposition time is maintained, the nominal amount of material from the QD layer increases from 4.5 ML in S1 to 7.5 ML in S5. As an example to visualize the structure, a high-angle annular dark field (HAADF) STEM image of sample S2 is displayed in figure 1b, together with a zoomed-in bright field (BF) STEM micrograph showing three QD planes, in figure 1c. The QD layers appear like rough quantum wells with in-plane changes of contrast associated with the presence of the QDs. This is due to the fact that the electron beam traverses more than one QD, since the QD diameter is significantly smaller than the thickness of the lamella (around 50 nm). From the fluctuations of contrast in STEM images, we can estimate a QD diameter of $6.2 \pm 1.3$ nm along the $\langle 1\bar{1}00 \rangle$ direction. The QD height is around $5.0 \pm 0.6$ ML, including the wetting layer, whose thickness is estimated at $2.5 \pm 0.5$ ML. The average values and error bars were extracted from the analysis of around 50 QDs in different sample regions. Compared to Stranski-Krastanov GaN or $Al_{0.1}Ga_{0.9}N$ QDs,[15,17,18] which present a typical aspect ratio (height/diameter) in the range of 0.13-0.20,[23,26,27] the QDs in figure 1c are flatter, with an aspect ratio around 0.10, which is explained by the reduced lattice misfit between the AlGaN dots and the AlN barrier, and the low mobility of Al at the growth temperature.

With respect to the superlattice period extracted from XRD (also presented in table 1), the trend is that higher $\Phi_{Al}$ results in a longer period, as expected. However, these data must be taken with caution, since the variations are within the error bars of the measurements, and in the same range as the thickness fluctuations that can be observed along the wafer.

Figure 2a shows the CL spectra of this series of samples at room temperature. Throughout the series, the emission wavelength shortens from 318 nm to 266 nm due to the Al incorporation, as summarized in table 1. In figure 2b, we compare the evolution of the CL peak emission wavelength with theoretical calculations. The "nominal" calculations (dotted line) assume that the QDs behave as quantum wells, with a thickness corresponding to the total amount of AlGaN deposited to form the QDs. Although this approximation assumes that the biaxial strain condition is valid, which is not usually the case in QD structures,[17,18] we consider it reasonable in this case, given the morphology of the QD layers in figure 1c. To assess the sensitivity to variations of the nominal growth parameters, the figure also presents the calculated emission wavelengths within reasonable variations in the QD thickness (±1 ML, dashed lines) and Al mole fraction (±10%, solid lines). The experimental data is located within the nominal calculation and those assuming that the QDs contain 1 ML less material than that of the nominal value. This might be explained either by the presence of a uniaxial stress along the growth direction, due to the 3D nature of the QDs, which would blue shift the emission,[17] or/and by the relatively high growth temperature and the tendency of AlN to etch away the GaN during the overgrowth process.[28] The apparent saturation of the emission wavelength observed experimentally for the two samples with higher Al content (S4 and S5) remains within the error bars marked by the calculations, indicating that it is probably due to a statistical deviation rather than a physical or growth-related limit on the emission wavelength.

The $IQE$, estimated from the CL integrated intensity ratio between room temperature and 5 K measurements in various regions of the sample, is also indicated in table 1. The series presents high *IQE* (36-72%), with values that are not correlated with the aluminum content, metal/N ratio or emission wavelength. The sample with the highest Al content, S5, displays an *IQE* of 51% at 266 nm, similar to that of the GaN sample, S1. This is noteworthy because it demonstrates that the degree of freedom to tune the metal/N ratio is large, allowing the growth of efficient QD emitters even at ratios as high as 0.8.

We have analyzed the homogeneity of the emission, classifying the samples within four categories. Depending on the relative deviation of internal quantum efficiency, the homogeneity is considered *very high* when $\Delta(IQE)/IQE < 12\%$, *high* when $\Delta(IQE)/IQE < 20\%$, *low* when $\Delta(IQE)/IQE < 40\%$ or *very low* when $\Delta(IQE)/IQE > 40\%$. All the samples from this series display *very high* homogeneity, except for S4, which is listed as having *high* homogeneity (see table 1).

The reduced emission homogeneity of the S4 sample is accompanied by a broadening of the emission towards longer wavelengths (see figure 2a). An SEM micrograph of the S4 sample surface is presented in figure 2c, revealing a high density of pits. Comparing this morphology with a λ-filtered CL map figure 2d, the emission at a longer wavelengths appears associated with the pits. Overall, the flat surface presents emission spectra that peak around 263 nm, whereas the emission in the proximity of the pits shifts to 280 nm, as shown in figure 2e.

**Series 2: Constant metal/N ratio**

To further reduce the emission wavelength, the second set of samples (S1, S6-S10) explores the influence of incorporating an aluminum flux, but now keeping a constant metal/N ratio (($\Phi_{Al} + \Phi_{Ga})/\Phi_N = 0.5$) during the growth of the $Al_xGa_{1-x}N$ QDs. The growth parameters, nominal AlGaN composition and structural and optical characterization data are summarized in table 2. Using the same values of $\Phi_{Al}$ as the first series, the reduction of $\Phi_{Ga}$ (varying from 0.3 to 0.1 ML/s) while keeping constant active nitrogen ($\Phi_N$ = 0.6 ML/s) allows reaching a nominal Al content of 65%.

Figure 3a presents the CL spectra of this series of samples at room temperature. The emission wavelengths decrease from 318 nm for the reference sample to around 230 nm for S8 and S9 due to the Al incorporation. However, we observe bimodal spectra for samples S8 and S9, with a prominent secondary contribution at longer emission wavelengths, around 265 nm. This is accompanied by an increase in emission inhomogeneity and the presence of pits on the surface of these samples.

The secondary spectral line is located in spectral region assigned to recombination involving nitrogen vacancies in AlN.[29,30] However, further insight into the nature of the longer-wavelength emission can be gained from the estimation of the *IQE* displayed in table 2. By considering only the primary emission peak at room temperature divided by the integrated CL intensity (full spectrum) at 5 K, the *IQE* tends to decrease for shorter emission wavelengths, with values in the range of 20-40% for peak emission wavelenths in the 227-240 nm spectral range. Nonetheless, the *IQE* of the whole emission remains around 50% throughout the series. This observation supports the hypothesis that the aforementioned longer-wavelength emission originates from a second population of QDs rather than from point defects, which would have led to a decrease of the radiative efficiency.

Figure 3b compares the evolution of the CL emission wavelength with theoretical calculations. The diagram shows only the short emission wavelength for samples with bimodal emission. Similar to the previous series, the experimental data falls within the range of both the nominal calculation and the calculation assuming that the QDs contain 1 ML less of material than estimated. As predicted by the theoretical model, the emission wavelength decreases as the Al mole fraction increases. However, the bimodality and increased inhomogeneity impose a constraint on the maximum aluminum concentration that can be efficiently incorporated in the QDs.

**Series 3: Reduced growth time**

We have attempted to further decrease the emission wavelength by reducing the QD growth time. In this regard, we have compared the Al-rich (x = 0.47 and x = 0.65) samples (S8-10) from series 2, which were grown with a QD deposition time of 15 s, with similar samples grown using a QD deposition time of 12 s (S11,S12). This implies reducing the total amount of AlGaN deposited to form the QDs from 4.5 ML in samples S8-10 to 3.6 ML in samples S11 and S12. Table 3 provides an overview of the growth parameters, nominal AlGaN composition as well as structural and optical characterization data.

The resulting CL spectra at 300 K are represented in figure 3c. Reducing the amount of deposited material should lead to a reduction of the QD size (height and/or diameter), since the density of QDs is mostly governed by the growth temperature that determines the capture radius around the QD nucleation sites. However, the comparison in figure 3c shows that this reduction of size does not come with a blue shift of the emission. To understand this result, we performed 3D calculations of the AlGaN QDs containing 47% of aluminum. We used a nominal structure that reproduces the S8 sample, with a QD diameter of 6.2 nm and a QD height of 1.17 nm over a wetting layer of 0.65 nm. The dimensions are extrapolated from the STEM measurements of S2 shown in figure 1c. Our simulations of the nominal structure at room teperature, taking the 3D strain distribution and exciton binding energy into account, predict a transition at 239.8 nm. The difference with the experimental value of 231 nm is within the error bars of our determination of the geometry and Al concentration.

Using the nominal calculations as a reference, if the QD height is reduced by 1 ML, the emission blue shifts around 6 nm. Conversely, if the QD diameter is decreased by 1.3 nm, which yields a similar volume reduction as a 1 ML decrease in QD height, the change has a smaller impact on the wavelength, resulting in a blue shift of about 2 nm. The structures considered for these calculations, as well as the extracted energy levels and transitions, are presented as Supporting Information. These results suggest that the emission wavelength of AlGaN QDs with high Al concentration shows lower sensitivity to changes in their diameter than to changes in their height. This finding contrasts with our previous calculations, which revealed a strong dependence of peak emission wavelength on the QD diameter for GaN/AlN and $Al_{0.1}Ga_{0.9}N$/AlN QDs due to the much larger larger lattice mismatch in these later cases.[18]

In conclusion, we have experimentally found that the reduction of growth time for AlGaN QDs with high Al concentration do not produce a significant blue shift of the emission. This can be attributed to the fact that the reduction in QD growth time leads mostly to a reduction of the QD diameter, while the QD height remains relatively constant.

On the other hand, samples with shorter QD deposition time present an enhancement of the secondary emission at longer wavelengths. This secondary line increases the spectral inhomogeneity in the emission over the surface of the samples, as shown in figure 3c. In spite of this inhomogeneity, the average $IQE$, considering the whole spectrum at room temperature, remains relatively high (30-50% range), which points again to an increased inhomogenetity of the QD geometry, rather than point defects.

**Discussion**

Figure 4a collects the $IQE$ data extracted from the three series of samples, and includes the samples from ref. [18] The different symbol shapes distinguish the series, and the color code indicates the nominal Al concentration in the QDs. All the samples exhibit $IQE$ values around 50%, with deviations that are not correlated with the metal/N ratio (which varies between 0.3 and 0.8), the Al content in the QDs (ranging from x = 0 to 0.65), or the emission wavelength (330 to 225 nm). However, the samples emitting at wavelengths shorter than 270 nm tend to display a bimodal emission, with a secondary long-wavelength peak that becomes more pronounced for shorter deposition time (resulting in less material in the QDs). This secondary line penalizes the emission at short wavelengths, but the $IQE$ of the whole spectrum remains similar to that of monomodal samples. Based on this information, we infer that the multiple lines are due to in-plane or layer-to-layer fluctuations or the QD shape, rather than point defects, which would enhance non-radiative recombination paths. The inhomogeneity of the CL maps (see figure 2d) suggests the presence of localized QD inhomogeneities associated with structural extended defects. Therefore, a detailed structural analysis is necessary to identify and correct the origin of the bimodality.

While examining the extension of the technology towards shorter wavelengths, it is also interesting to compare the spectral behavior of the samples at low temperature and at room temperature. The semiconductor band gap decreases with temperature following a trend that is modelled by Varshni empirical equation. The expected red shift for AlGaN is represented as a dashed line in figure 4b as a function of the Al mole fraction. Using ternary alloys, it is common for the red shift to be smaller than the corresponding theoretical value. At low temperature, alloy fluctuations can result in the localization of carriers in areas with smaller band gap. Then, increasing the temperature, carriers get enough energy to overcome the localization, which leads to a blue shift of the emission, which partially compensates Varshni's red shift.

Examining the experimental results shown in figure 4b, which compares the spectral shift between low and high temperatures for series 1, 2 and 3 to the theoretical trend, it becomes apparent that the red shift is significantly higher than the expected value in samples with higher Al concentrations. This trend cannot be attributed to alloy fluctuations. However, it appears associated with an increase in inhomogeneity of the in-plane emission and the onset of bimodality. Despite this, the $IQE$ of theses samples remains high and comparable to homogeneous samples emitting at longer wavelengths, suggesting a thermally-induced transfer of carriers between QDs with different sizes. At low temperature, carriers are localized in the dots; but as the temperature increases, the in-plane carrier mobility is enhanced, particularly via the large wetting layers, which enables the migration of carriers towards bigger QDs that emit at longer wavelengths. This process, favored by a lower band offset with respect to the samples with low Al content, results in a more pronounced red shift than Varshni's prediction.

Finally, the power efficiency of the samples was extracted from measurements of the angular distribution of the UV emission under electron beam pumping. For this purpose, a calibrated GaP photodetector was moved around the sample as indicated in figure 5a. An example of angular measurement is presented in arbitrary units in figure 5b. From these data, the optical power emitted by the sample can be calculated as:

$$P_{opt} = \frac{1}{R_\lambda} \int_0^{\pi/2} \frac{I_d(\theta)}{A_d} 2\pi R^2 \sin\theta \, d\theta \quad (1)$$

where $R_\lambda$ is the responsivity of the photodiode at the emission wavelength, R is the distance between the photodiode and the sample, $A_d$ is the area of the photodiode, $I_d$ is the photocurrent, and $\theta$ is the angle of rotation around the sample. For these measurements, the as-grown samples were mounted with silver paste on a copper plate. The surfaces were not treated in any way (neither roughening nor metallization) to improve the light extraction. The extracted values of power efficiency, summarized in tables 1 and 2, remain in the range of 0.4-1% for most of the samples. These relatively low values with regard to the $IQE$ can be attributed to the low light extraction efficiency. Notably, we did not observe a clear correlation between emission wavelength or Al content and power efficiency. This is noteworthy, as it suggests that highly efficient (power efficiency higher than 10%) far UV-C emitters for wavelengths as short as 220 nm could be produced if the light extraction efficiency is improved.

Figure 5c compares the power efficiency values from AlGaN/AlN QD-SLs with data in the literature for electron beam pumped samples containing AlGaN-based planar materials (AlGaN layers or quantum wells),[9–12,31–35] and hexagonal BN[36]. The values of power efficiency are higher than those of LEDs at short wavelengths (below 250 nm), but they are lower than those of UV-emitting phosphors (e.g. $Lu_2Si_2O_7$:$Pr^{3+}$, $YPO_4$:$Bi^{3+}$, YPO4:$Pr3^+$, Y2SiO5:$Pr^{3+}$, YBO3:$Pr3^+$ or LuAG:Pr)[37–40], which can display efficiencies in the range of 1-9% (estimations from a point measurement assuming that phosphors are Lambertian emitters)[40]. The phosphor morphology, typically consisting of micrometer-sized grains, intrinsically favors the light extraction. However, Pr-based phosphors present a multiband emission in the range of 225-350 nm, whereas Bi-based materials present a single-band emission peaking at 240-250 nm, but $Bi^{3+}$ cations are highly sensitive to oxidation or reduction.

In the case of nitride semiconductors, the top priority is to enhance the light extraction efficiency. In LEDs, this goal is often achieved through a combination of flip-chip bonding and back-side roughening or patterning techniques.[41–44] In spite of these efforts, the light extraction efficiency remains limited to 6-8% due to absorption in the p-type layers.[5] In the case of electron beam pumped devices, p-type layers are not necessary. However, integrating an Al-based mount as a reflector can be crucial, and the use of patterned sapphire substrates promises an important improvement. For electron beam pumped lamps with transmission geometry, encapsulation with a hemispherical silicone dome could also be a viable option to improve light extraction.[45]

Another important consideration is thermal management. Several experiments, including those conducted by Ivanov et al. (up to 60 mW emission)[11], and Jmerik et al. (up to 150 mW)[34], and by Rong et al. (up to 2.2 W)[35], have shown a significant improvement of the efficiency (by almost a factor of 2) under pulsed excitation. This suggests that the sample mounting may not be sufficient for effective heat evacuation, and therefore a redesign may be necessary to address this issue. Implementing such improvements would also benefit the results described in this paper.

## 4. CONCLUSION

The present study investigates the synthesis and performance of highly efficient $Al_xGa_{x-1}N/AlN$ QD superlattices for electron beam pumped emitters operating in the far UV-C range. The target is to reduce the emission wavelength below 240 nm, for the fabrication of UV-C lamps to enable UV disinfection of occupied spaces. Achieving such highly energetic emission requires producing extremely small QDs with high Al content, which implies reduced lattice mismatch and band offset with the AlN matrix. By using the Stransky-Krastanov growth mode in plasma-assisted molecular beam epitaxy, we demonstrate here QD samples with internal quantum efficiency around 50%, with deviations that are not correlated with the Al content in the QDs (up to 65%) or the emission wavelength (in the 300-230 nm range). The degree of freedom to adjust the metal/N ratio during the QD deposition is large, allowing the growth of efficient QD emitters even at ratios as high as 0.8. However, samples emitting at wavelengths below 270 nm displayed a bimodal luminescence associated with inhomogeneous in-plane emission and higher-than-expected thermal red shift that hindered the shift of the emission towards shorter wavelengths. The bimodality was attributed to fluctuations of the QD shape associated to extended defects. Reducing the QD growth time (i.e. deposition of less than 4

ML of AlGaN to form the QDs) exacerbated the bimodality without reducing the emission wavelength, since the effect of reducing the amount of material is higher on the QD diameter than on the QD height. The power efficiencies of the samples under electron beam pumping were in the range of 0.4-1%, without clear correlation with the emission wavelength or Al content. The efficiencies are higher than those demonstrated by AlGaN LEDs in the 240-220 nm range, and there is still room for improvement through the enhancement of the light extraction efficiency. Overall, this work is a promising further step towards the development of efficient UV-C sources and it provides new valuable insights for the development of AlGaN-based electron-pumped emitters.

## ■ SUPPORTING INFORMATION

XRD θ-2θ scan along the (0002) reflection of AlN of the S1 sample and structures, parameters and results from Nextnano3 simulations reproducing the S8 sample.

## ■ ACKNOWLEDGEMENTS

This project received funding from the French National Research Agency (ANR) via the ASCESE-3D and FUSSL projects (ANR-21-CE50-0016 and ANR-22-CE09-0024), and by the Auvergne-Rhône-Alpes region (PEAPLE grant). A CC-BY public copyright license has been applied by the authors to the present document and will be applied to all subsequent versions up to the Author Accepted Manuscript arising from this submission, in accordance with the grant's open access conditions.

Part of this work, performed on the Platform for NanoCharacterisation (PFNC) of CEA, was supported by the "Recherche Technologique de Base" Program of the French Ministry of Research.

## ■ REFERENCES

**Figure 1**

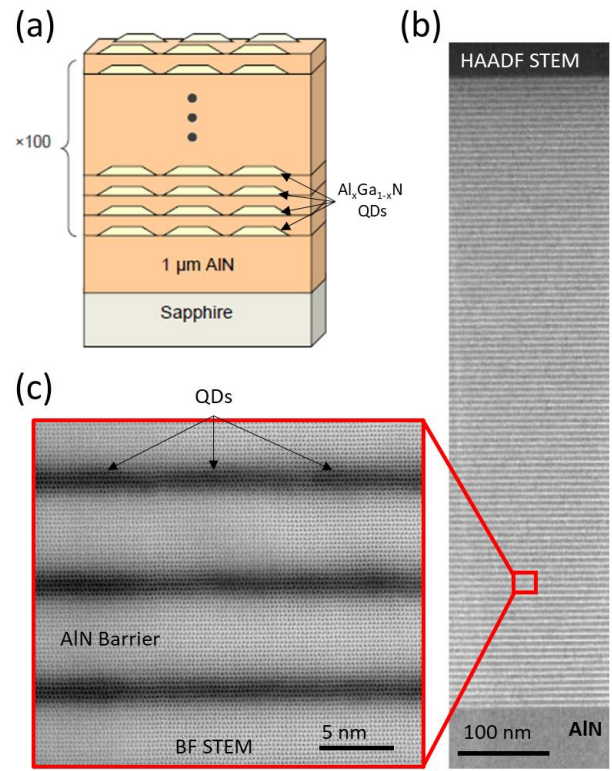

**Figure 1.** (a) Sample structure consisting of 100 periods of $Al_xGa_{1-x}N$ QDs embedded in AlN barriers. (b) HAADF-STEM image of the full QDs SL in the sample S2. (c) BF-STEM zoomed-in micrograph showing three QD planes.

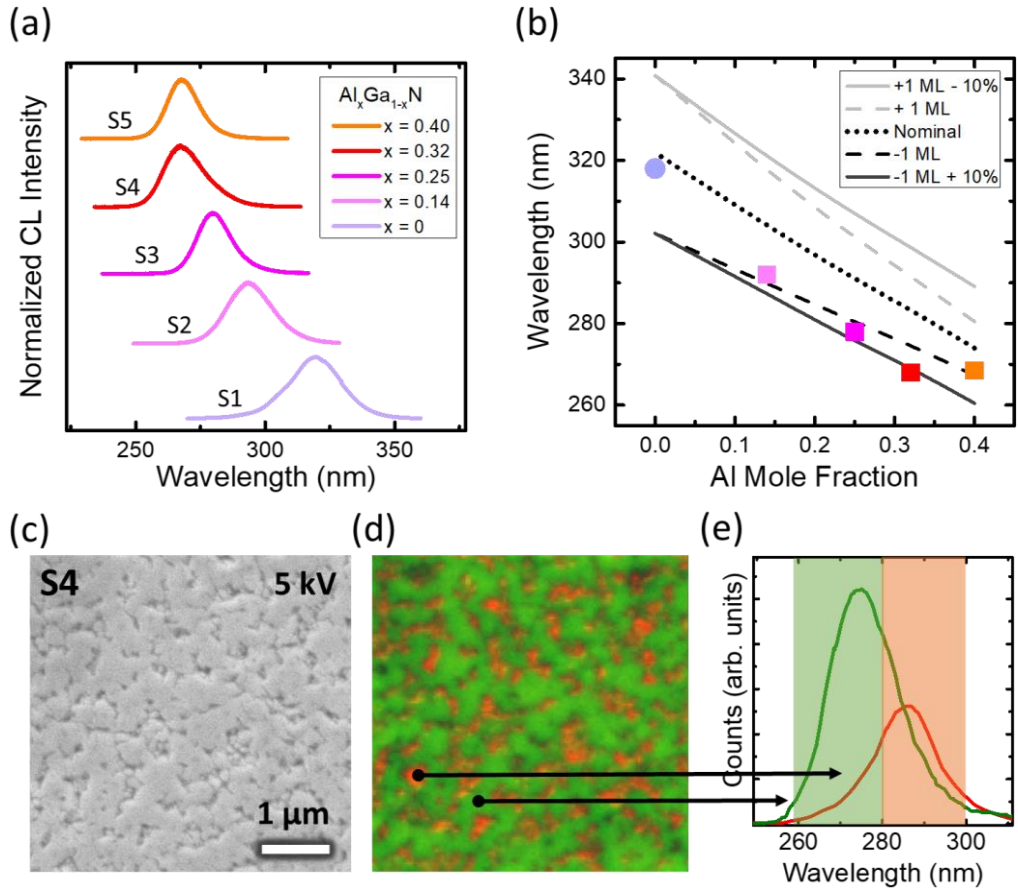

**Figure 2.** (a) Room-temperature CL emission from series 1 (samples with QDs grown with constant gallium flux). (b) Experimental versus calculated emission wavelengths for series 1. The dots represent the experimental points. The calculated emissions errors are estimated by calculating the emission within reasonable variations in the QDs thicknesses and compositions of ±1 ML and ±10% of the Al mole fraction. (c) Top-view SEM image and (d) λ-filtered CL map of S4. (e) Local emission spectra from the vicinity of the pits (red) and from the flat surface (green).

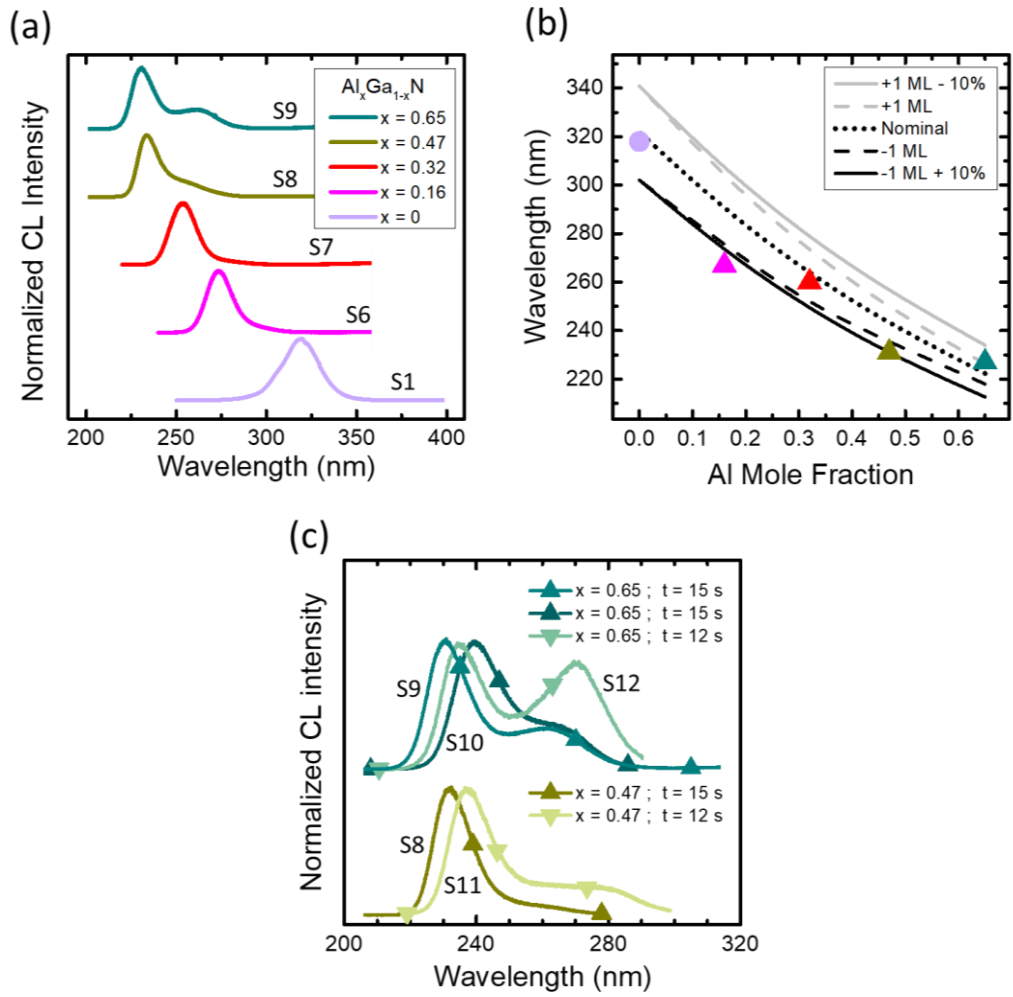

**Figure 3.** (a) Room-temperature CL emission from series 2 (QDs grown with constant metal/N ratio). (b) Experimental versus calculated emission wavelengths for series 2. (c) Room-temperature CL emission from the samples of series 3 (reduction of the growth time).

**Figure 4**

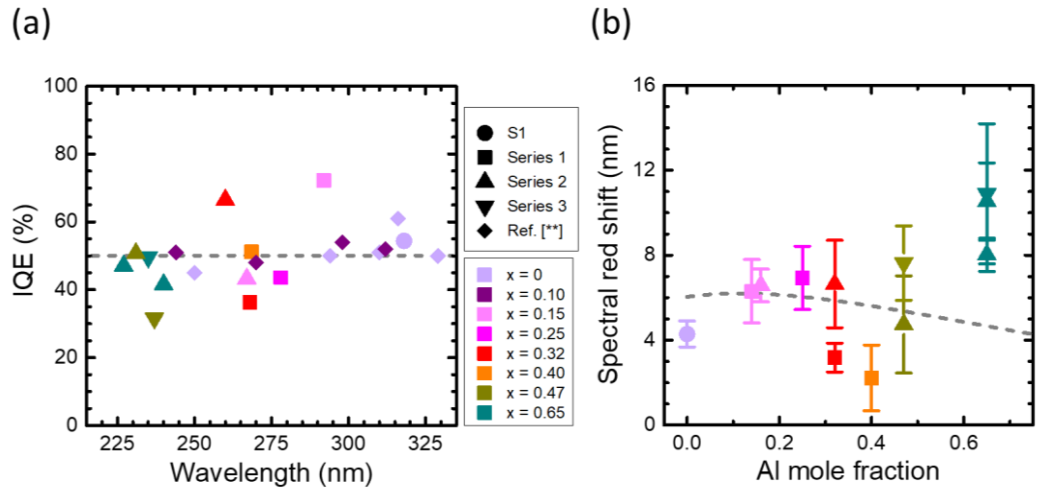

**Figure 4.** (a) Comparison of the *IQE* of all samples under study, considering the whole spectrum at room temperature. For completion, we have added the results from ref. [18]. (b) Spectral shift between the low-temperature and room-temperature emission for all samples under study. The dashed line describes the expected values from Varshni equation.

**Figure 5**

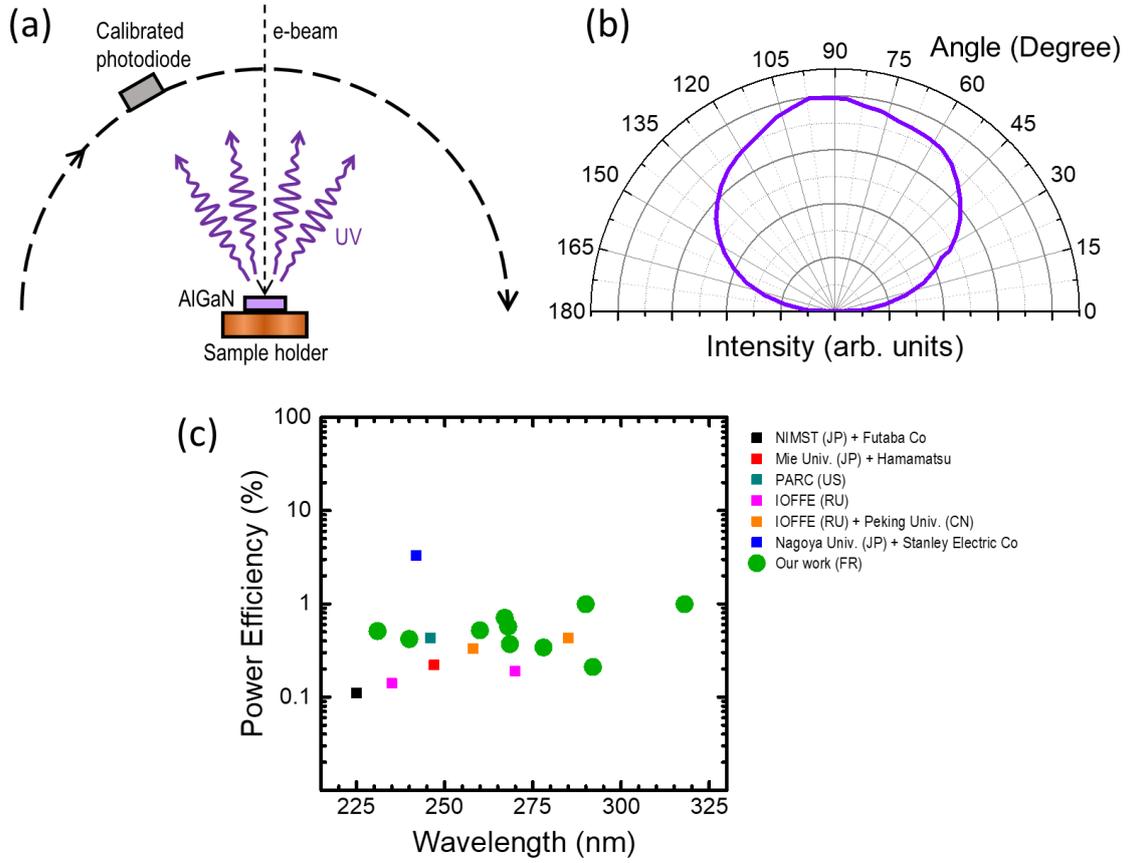

**Figure 5.** (a) Schematic description of the setup to measure the power efficiency under electron beam injection. (b) Angular distribution of the emission intensity (sample S8). (c) Power efficiency represented as a function of the peak emission wavelength. Our results are compared with nitride-based electron beem pumped samples in the literature, all measured under continuous excitation.

# Tables

| Sample | x | $\Phi_{Ga}$ (ML/s) | $\Phi_{Al}$ (ML/s) | Metal/N | Deposition time (s) | Period (nm) | $\lambda_{LT}$ (nm) | $\lambda_{RT}$ (nm) | IQE | Homogeneity | PE |
|---|---|---|---|---|---|---|---|---|---|---|---|
| S1 | 0 | 0.3 | 0.000 | 0.500 | 15 | 5.50 | 314 | 318 | 54% | Very high | 0.99% |
| S2 | 0.14 | 0.3 | 0.049 | 0.581 | 15 | 5.71 | 287 | 292 | 72% | Very high | 0.21% |
| S3 | 0.25 | 0.3 | 0.099 | 0.666 | 15 | 6.05 | 273 | 278 | 44% | Very high | 0.34% |
| S4 | 0.32 | 0.3 | 0.142 | 0.737 | 15 | 5.80 | **263**, 280 | 268 | 36% | High | 0.57% |
| S5 | 0.40 | 0.3 | 0.198 | 0.831 | 15 | 6.18 | 266 | 268.5 | 51% | Very high | 0.37% |

**Table 1.** Growth conditions and various structural and optical characterization values for series 1: Al concentration in the QDs (x), Ga and Al fluxes ($\Phi_{Ga}$ and $\Phi_{Al}$, respectively), metal/N ratio, QD deposition time, period extracted from XRD, low ($\lambda_{LT}$) and high ($\lambda_{LT}$) temperature peak emission wavelength, internal quantum efficiency (IQE), homogeneity of the emission, and power efficiency (PE). In the case of multiple emission peaks, the main peak emission wavelength appears in bold.

| Sample | x | $\Phi_{Ga}$ (ML/s) | $\Phi_{Al}$ (ML/s) | Metal/N | Deposition time (s) | Period (nm) | $\lambda_{LT}$ (nm) | $\lambda_{RT}$ (nm) | IQE | Homogeneity | PE |
|---|---|---|---|---|---|---|---|---|---|---|---|
| S1 | 0 | 0.3 | 0 | 0.5 | 15 | 5.50 | 314 | 318 | 54% | Very high | 0.99% |
| S6 | 0.16 | 0.258 | 0.049 | 0.511 | 15 | 5.55 | **261**, 290 | 267 | 43% | High | 0.71% |
| S7 | 0.32 | 0.212 | 0.099 | 0.519 | 15 | 5.60 | **254**, 280 | **260**, 280 | 67% (60%) | High | 0.52% |
| S8 | 0.47 | 0.162 | 0.142 | 0.507 | 15 | 5.55 | **227**, 255, 272 | **231**, 255 | 51% (39%) | Low | 0.51% |
| S9 | 0.65 | 0.107 | 0.198 | 0.509 | 15 | 5.71 | **219**, 263 | **227**, 265 | 47% (17%) | Low | 0.10% |
| S10 | 0.65 | 0.107 | 0.198 | 0.509 | 15 | 5.47 | **225**, 254 | **240**, 265 | 42% (28%) | Low | 0.42% |

**Table 2.** Growth conditions and various structural and optical characterization values for series 2: Al concentration in the QDs (x), Ga and Al fluxes ($\Phi_{Ga}$ and $\Phi_{Al}$, respectively), metal/N ratio, QD deposition time, period extracted from XRD, low ($\lambda_{LT}$) and high ($\lambda_{LT}$) temperature peak emission wavelength, internal quantum efficiency (IQE), homogeneity of the emission, and power efficiency (PE). In the case of multiple emission peaks, the main peak emission wavelength appears in bold. In the case of multiple emission peaks at room temperature, the table contains the IQE value considering the whole spectrum and, between parenthesis, the IQE value considering only the high energy peak at room temperature.

| Sample | x | $\Phi_{Ga}$ (ML/s) | $\Phi_{Al}$ (ML/s) | Metal/N | Deposition time (s) | Period (nm) | $\lambda_{LT}$ (nm) | $\lambda_{RT}$ (nm) | IQE | Homogeneity |
|---|---|---|---|---|---|---|---|---|---|---|
| S8 | 0.47 | 0.162 | 0.142 | 0.507 | 15 | 5.55 | **227**, 255, 272 | **231**, 255 | 51% (39%) | Low |
| S9 | 0.65 | 0.107 | 0.198 | 0.509 | 15 | 5.71 | **219**, 263 | **227**, 265 | 47% (17%) | Low |
| S10 | 0.65 | 0.107 | 0.198 | 0.509 | 15 | 5.47 | **225**, 254 | **240**, 265 | 42% (28%) | Low |
| S11 | 0.47 | 0.161 | 0.142 | 0.506 | 12 | 5.35 | **228**, 274 | **237**, 277 | 32% (20%) | Very low |
| S12 | 0.65 | 0.107 | 0.198 | 0.509 | 12 | 5.65 | **223**, 263 | **235**, 265 | 50% (31%) | Very low |

**Table 3.** Growth conditions and various structural and optical characterization values for series 2: Al concentration in the QDs (x), Ga and Al fluxes ($\Phi_{Ga}$ and $\Phi_{Al}$, respectively), metal/N ratio, QD deposition time, period extracted from XRD, low ($\lambda_{LT}$) and high ($\lambda_{LT}$) temperature peak emission wavelength, internal quantum efficiency (IQE), homogeneity of the emission, and power efficiency (PE). In the case of multiple emission peaks, the main peak emission wavelength appears in bold. In the case of multiple emission peaks at room temperature, the table contains the IQE value considering the whole spectrum and, between parenthesis, the IQE value considering only the high energy peak at room temperature.

# Supporting information

**AlGaN/AlN Stranski-Krastanov quantum dots for highly efficient electron beam pumped emitters: The role of miniaturization and composition to attain far UV-C emission**

Jesus Cañas[*,1], Anjali Harikumar[1], Stephen T. Purcell[3], Nevine Rochat[2], Adeline Grenier[2], Audrey Jannaud[2], Edith Bellet-Amalric[1], Fabrice Donatini[4], and Eva Monroy[1]

[1] Univ. Grenoble-Alpes, CEA, Grenoble INP, IRIG, PHELIQS, 38000 Grenoble, France
[2] Univ. Grenoble Alpes, CEA, LETI, 38000 Grenoble, France
[3] Institut Lumière Matière, CNRS, University of Lyon, Univ. Claude Bernard Lyon 1, 69622 Villeurbanne, France
[4] Univ. Grenoble Alpes, CNRS, Grenoble INP, Institut Neel, 38000 Grenoble, France

* Corresponding author: jesus.canasfernandez@cea.fr

## X-ray diffractogram of the S1 sample

The deposition of pure GaN during 15 s (4.5 ML) leads to the spontaneous formation of quantum dots (QDs) on AlN. A θ–2θ x-ray diffraction (XRD) scan around the (0002) reflection of AlN in the reference sample, S1, is presented in Figure S1. A superlattice (barrier + QDs) period of 5.5 nm is deduced from the 2θ spacing of the satellite peaks of the superlattice.

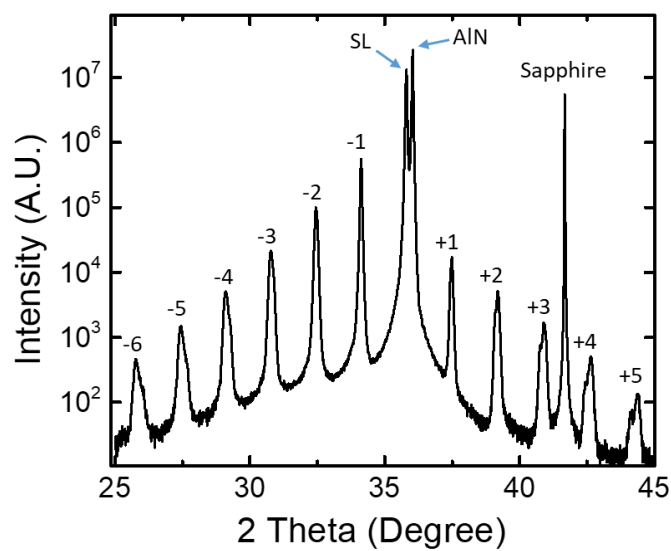

**Figure S1.** XRD θ-2θ scan around the (0002) reflection of AlN. Labels indicate the (0002) reflection of the AlN and the superlattice (SL) with several satellites. The peak at 41.7 degrees correspond to the (0006) reflection of the sapphire substrate.

**Nextnano calculations**

Three-dimensional calculations for the case of AlGaN QDs containing 47% of aluminum were performed solving the Poisson and Schrodinger equations with Nextnano3. We used a nominal structure that reproduces the S8 sample, with a QD height of 1.17 nm including the wetting layer and a QD diameter (along the $\langle 11\bar{2}0 \rangle$ direction) of 6.2 nm. According to our simulations, the nominal structure produces an emission at 239.8 nm. If the QD height is reduced by 1 ML, the emission shifts significantly down to 233.9 nm. Conversely, reducing the QD diameter by 1.3 nm without altering their density, which yields a similar volume reduction as a 1 ML decrease in QD height, results in a blue shift of 2 nm. The structural parameters of the structures used for the simulations and the resulting energy levels and transition wavelengths extracted from the calculations are presented in Table S1. In addition, a representation of a QD plane of the three simulated structures projected along the $\langle 1\bar{1}00 \rangle$ direction is presented in Figure S2. The squared wavefuctions of heavy holes and electrons are also displayed for each structure.

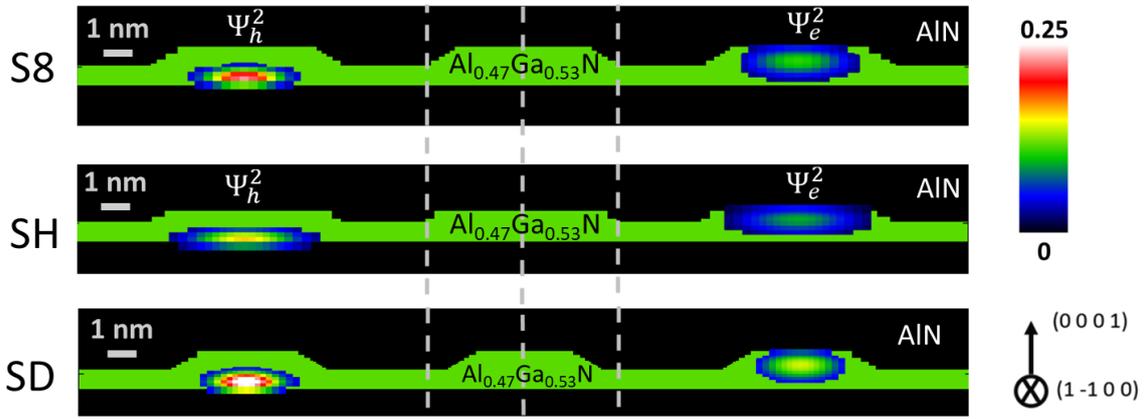

**Figure S2.** Representation of a QD plane of the simulated structures projected along the $\langle 1\bar{1}00 \rangle$ direction. The squared wavefuctions of heavy holes and electrons are also displayed for each structure. S8 describes the nominal structure of the grown sample. SH is the same structure as S8 but with a reduction of 1 ML in the QD height. SD is the same structure as S8 but with a reduction of 1.3 nm in the QD diameter.

|  | QD height (nm) | WL height (nm) | Total height (nm) | Diameter (nm) | e-h (eV) | Exciton (eV) | Energy (eV) | Wavelength (nm) |
|---|---|---|---|---|---|---|---|---|
| **Nominal (S8)** | 0.52 | 0.65 | 1.17 | 6.2 | 5.24 | 0.07 | 5.17 | 239.8 |
| **Smaller height (SH)** | 0.26 | 0.65 | 0.91 | 6.2 | 5.38 | 0.08 | 5.30 | 233.9 |
| **Smaller diameter (SD)** | 0.52 | 0.65 | 1.17 | 4.9 | 5.29 | 0.07 | 5.22 | 237.7 |

**Table S1.** Structural parameters used for the simulations: QDs height, wetting layer (WL) height, total height, and QD diameter. The energy distance between the first electron and hole levels (e-h), exciton energy, and optical transition energy and wavelength extracted from the calculations are also displayed in the table.